\documentclass[10pt,twocolumn,letterpaper]{article}

\usepackage{cvpr}              % To produce the CAMERA-READY version
\usepackage{graphicx}
\usepackage{amsmath}
\usepackage{amssymb}
\usepackage{booktabs}
\usepackage{cite}
\usepackage{amsfonts}
\usepackage[pagebackref,breaklinks,colorlinks]{hyperref}
\graphicspath{{figures/}}
\usepackage{float}
\def\cvprPaperID{*****} % *** Enter the CVPR Paper ID here
\def\confName{CVPR}
\def\confYear{2022}

\begin{document}

\title{Attention based Broadly Self-guided  Network for Low light Image Enhancement}
\author{Zilong Chen \\South China University of \\Technology  \and Yaling Liang \\South China University of \\ Technology \and Minghui Du \\South China University of \\ Technology}
% \\
% \and
\maketitle
\begin{abstract}
 Recently, Deep Learning have achieved impressive breakthroughs in low-light Image Enhancement. Existing deep learning methods mostly enhance the ability of feature extraction by stacking network structures and deepening the depth of the network. which causes more runtime cost on single image. In order to reduce inference time while fully extracting local features and global features. Inspired by SGN~\cite{gu2019self}, we propose a Attention based Broadly Self-guided Network (ABSGN) for real world low-light image Enhancement. The basic structure of ABSGN is a top-down self-guidance architecture which is able to efﬁciently incorporate multi-scale information and extract good local features to recover clean images. Moreover, such a structure requires a smaller number of parameters and enables us to achieve better effectiveness than U-net structure. In addition, The proposed network comprises several Multi-level guided Dense Blocks (MGDB), which can be viewed as a novel extension of Dense block in feature space. At the lowest resolution level of ABSGN, we offer more effecitenly module to fully extract the global information to generate the better final output, which called Global Spatial Attention (GSA). such a broadly strategy is able to handle the noise at different exposures. The proposed network is validated by many mainstream benchmark. Additional experimental results show that the proposed network outperforms most of state-of-the-art low-light image Enhancement solutions.
\end{abstract}
\section{Introduction}
\begin{figure}[t]
 \centering
  \begin{subfigure}{0.33\linewidth} 
  \begin{minipage}[t]{0.33\linewidth}  
	\includegraphics[width=1.0in]{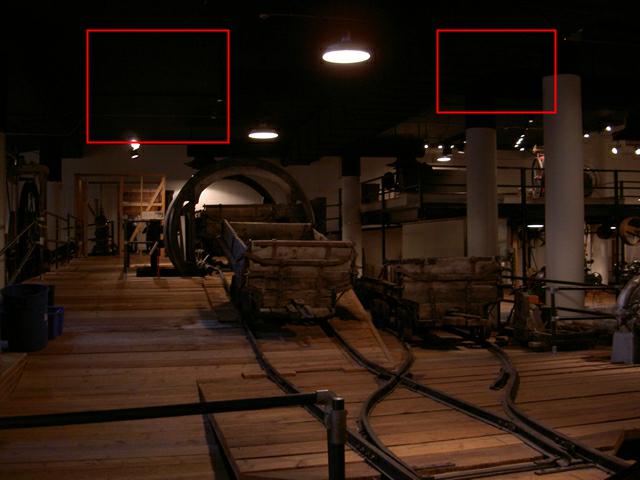}\vspace{0.1cm}
	\includegraphics[width=1.0in]{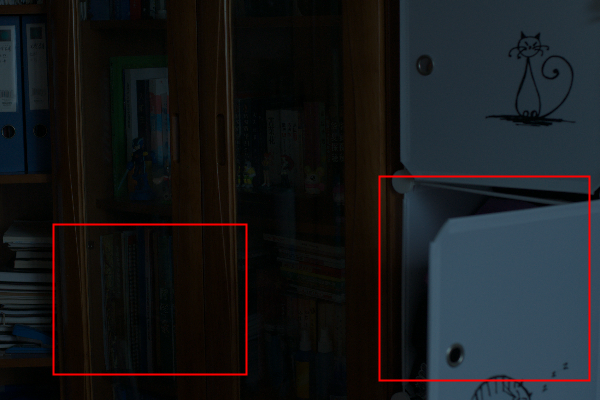}
 	\end{minipage}
 \caption*{Input}
\end{subfigure}\hspace{-1mm}
\begin{subfigure}{0.33\linewidth}
\begin{minipage}[t]{0.33\linewidth}  
\includegraphics[width=1.0in]{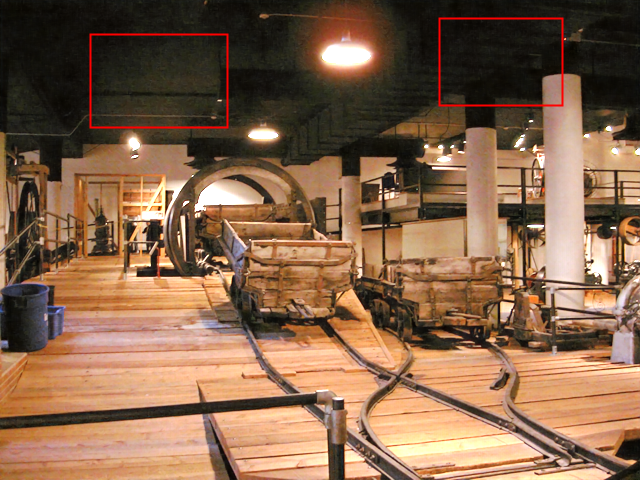}\vspace{0.1cm}
\includegraphics[width=1.0in]{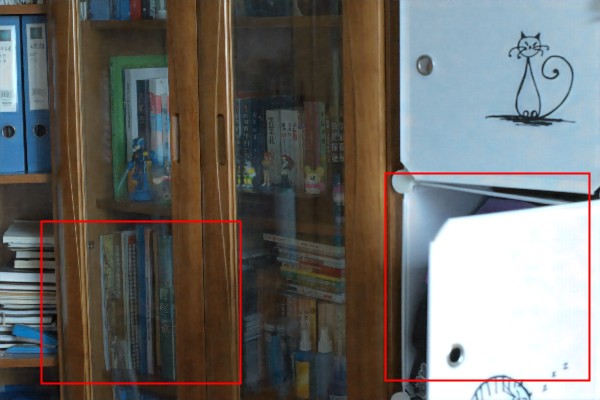}
\end{minipage}
\caption*{MIRNet}
\end{subfigure}\hspace{-1mm}
\begin{subfigure}{0.33\linewidth}
\begin{minipage}[t]{0.33\linewidth} 
\includegraphics[width=1.0in]{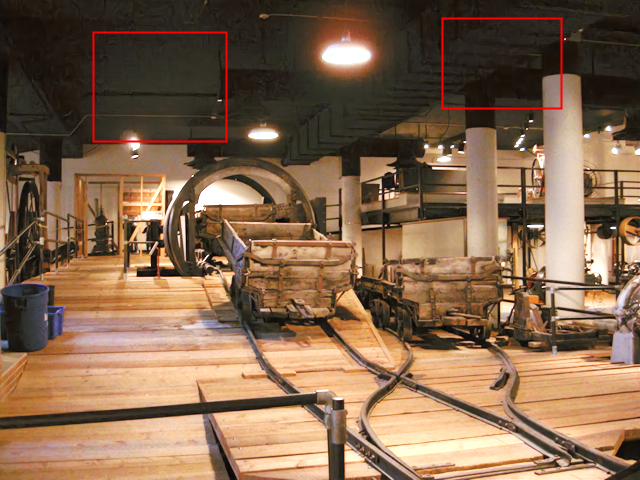}\vspace{0.1cm}
\includegraphics[width=1.0in]{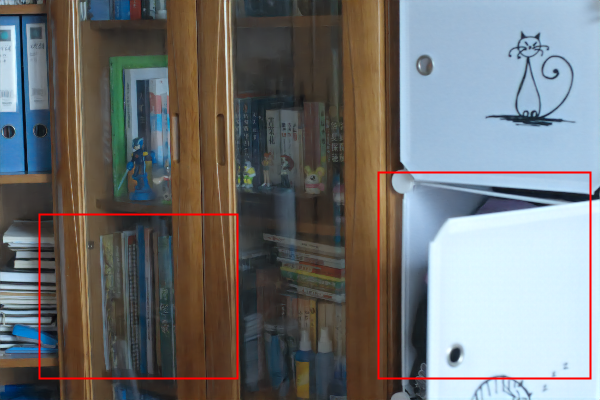}
\end{minipage}
\caption*{Ours}
\end{subfigure}
\caption{Visual comparision with supervised low light image enhancement method MIRNet~\cite{han2020mirnet}.The proposed method can well imporve the contrast and reduce noises and at same time reduce the color bias.}
\vspace{-0.5cm}
\end{figure}
Low-light Image Enhancement is a fundamental task in low-level vision and an important pre-processing step in many other vision tasks. images captured under the unsuitable lighting are either too dark or too bright. The art of recovering the original clean image from its corrupted measurements is studied under the image restoration task. It is an ill-posed inverse problem,due to the existence of many possible solutions. While we enhancing the brightness, we also need to tackle the color distortion, the amplified noise, the loss of detail and texture information and the blurred edges. Traditional methods usually address low-light image enhancement by histogram equalization(HE)-based approaches~\cite{pizer1987adaptive,pizer1990contrast,2007Brightness}, Retinex Theory~\cite{1978The,jobson1997properties,Jobson1997A}, Gamma Correction~\cite{rahman2016adaptive}, etc. However, these methods need to manually set parameters,have poor generalization ability and often result in visible noise for real low-light images. With the rapid development of deep learning technology, numerous advanced approaches~\cite{2017LLNet,2018LLCNN} have been developed and achieve impressive success. By involving the strategy of a noise-robust autoencoder-based way and ReLU activate function into deep architectures, LLNet~\cite{2017LLNet} is proposed to enlighten images with minimum pixel-level saturation and achieves a much higher peak signal (PSNR) to noise ratio than conventional state-of-the-art approaches~\cite{guo2016lime}. For the pursuit of highly accurate restoring results,some follow-up works have been proposed to decomposes low-light input into reflectance and illumination maps~\cite{wei2018deep,zhang2019kindling}. To involve multi-scale information,some most advanced end-to-end methods~\cite{2020Attention,han2020mirnet} apply U-net~\cite{ronneberger2015u} as their basic structure and add some dense residual block in each level. Although these methods obtain competitive performance on benchmark datasets, their heavy inference time hinder their application.

To seek a better trade-off between image denosing performance and the consumption of computational resources, Densely self-guided Wavelet network (DSWN)~\cite{2020Densely} is proposed for image denoising task by a top-down guidance strategy. DSWN generates multi-resolution inputs with the discrete wavelet  transformation (DWT) and inverse discrete wavelet transform (IDWT) before any convolutional operation. Large-scale contextual information extracted at low resolution is gradually propagated into the higher resolution sub-networks to guide the feature extraction processes at these scales. Using such a structure,DSWN is able to achieve a better denoising performance than U-net with less runtime and GPU memory.

Inspired by DSWN, we proposed a  Attention based Broadly Self-guided Network (ABSGN) (as is shown in Figure 2) which is able to improve performance of MIRNet (as is shown in Figure 1) and require less runtime than the state-of-the-art densely networks based on U-net structure. In ABSGN, we design Global Spatial Attention (GSA) Block (as shown in Figure 3) to better get the global information at the lowest resolution level. Then, we embed dilation convolution into Densely conneted Block to enlarge receptive Field using the self-guided strategy~\cite{gu2019self}, which we called Multi-level Guided Dense Block (MGDB), To achieve a better performance, we adopt more MGDB blocks with dense connections at the full resolution level. We combine discrete wavelet transformation (DWT) and a convolution layer with 3x3 convolution kernel to achieve the upsampling process, Corresponding to which we use inverse discrete wavelet transform (IDWT) and a convolution layer with 3x3 convolution kernel to achieve the downsampling process. Such a design can better extract the information of multi-scale feature maps. In addition,wavelet has been applied to denoising task in traditional methods\cite{sardy2001robust}. Utilizing wavelet transform to incorporate multi-scale information makes it possible for the network to have time-frequency analysis capabilities. Such a structure is able to help our network to deal with the low-light image enhancement task at different exposures.Our main contributions are concluded as follows:
\begin{itemize}
    \item As we konw,we firstly introduce Self-guided Network and Multi-level wavelet transform into low-light image enhancement,which can accelerate the inference speed and avoid down-sampling information loss.
    \item We embed dilation convolution into Dense Block to enlarge the receptive Field using the self-guided strategy,which achieve a higher PSNR and preserve more details.
    \item We propose a new global Attention module called GSA,which has better global feature extraction capability and further improve image restoration performance.
    \item We design a broadly self-guided wavelet network which outperforms conventional methods and is more efﬁcient than the state-of-the-art low-light image enhancement networks with dense blocks.
\end{itemize}

\section{Related works}
In this section, we briefly introduce some works related to our research. First, we review some deep learning based low-light image Enhancement, then we discuss some previous works including Attention mechanism and Feature extraction.
\subsection{Deep Neural Networks for low-light image enhancement}
In recent years, researches have shown that deep learning technologies outperform traditional methods on low-light image Enhancement by extracting more suitable image features~\cite{2018GLADNet}. including supervised learning,unsupervised learning and zero-shot learning. Most of the early works based on supervised learning train the networks. As we know,K.G.Lore et.al~\cite{2017LLNet} firstly designed a deep neural networks with stacked sparse denosing autoencoder perform low-light image enhancement with minimum pixel-level saturation.Using feedforward CNN with different Gaussian convolution kernels, MSR-Net~\cite{shen2017msr} is able to simulate the pipeline of Multi-scale Retinex (MSR) for directly learning end-to-end mapping between dark and bright images. GLADNet~\cite{2018GLADNet} is able to calculates global illumination and has global illumination-aware, which better handle a wide range of light levels.  By introducing Retinex theory to explicitly decompose the image into reflectance and illumination, RetinexNet~\cite{wei2018deep} is able to enhance the lightnesss over illumination. The similar work KinD~\cite{zhang2019kindling} additionally introduce degradation removal in the reflectance to improve the quality of the restored image. KinD++ is the improved version of KinD,which Effectively improve the quality of reflection map by using Multi-scale illumination attention(MISA) module. Considering both Retinex model and GAN, RDGAN~\cite{2019RDGAN} further improved the restored image quality by embedding a GAN network.

Besides, by involving the strategy of generative adversarial network (GAN) into deep architectures, EnlightenGan~\cite{2021EnlightenGAN} is proposed to handle the limitation of low/normal-light pairs, which learns a one-to-many relation from low-light to normal-light image space without paired datasets. To overcome the lack of paired training data, Zhang et.al ~\cite{zhang2021self} presented a self-supervised low-light Image Enhancemnet, which fully make use of Retinex theroy to decompose the low-light image and enhance the reflectance map to view as the final restored image. In addition, Zero-DCE~\cite{2020Zero} estimates the pixel-wise and the high-order curves for dynamic range adjustment of a low-light image in an unsupervised way, which use meticulously designed non-reference loss functions for training. Recently, Syed Waqas Zamir et al~\cite{han2020mirnet} proposed a Multiscale Residual Block (MRB) that uses both residual learning and Attention Unit as its basic structure, maximizing feature reuse and achieving a significant improvement in the performance of low light image enhancement.
\subsection{Attention mechanism}
The attention mechanism has been studied in a wide variety of computer vision problems, which try to mimic the human visual system in making substantial use of contextual information in understanding RGB images. This involves discarding unwanted regions in the image, while focusing on more important parts containing rich features of our vision task. Based on such an idea,SENet~\cite{hu2018squeeze} is to learn feature weights according to loss through the network,so that the effective feature map weight is large, and the invalid or small effect is small. Using such a strategy to train the model to achieve better results. Further, Woo,S.et.al propose Channel Attention Module (CBA) and Spatial Attention module (SPA) to pay attention to different components. In this paper,we propose a novel Global attention CNN model. In which,We use the spatial Attention to improve the attention to different area in feature maps. Since the feature map of each channel has different contributions to the following network. Therefore, we also embed CBA into our MGDB to extract more useful information for low-light image enhancement. The structure of channel attention is illustrated in Figure 2.
% \subsection{Extraction of multi-scale information}
% To extract multi-scale information for image restoration tasks,With self-guidance strategy and PixelShuffle,SGN~\cite{gu2019self} greatly improved the memory and runtime efﬁciency. Bae et.al proposed a wavelet residual network(WavResNet)~\cite{2017Beyond} for image denoising and SISR and ﬁnd wavelet subbands benefits learning convolutional neural network (CNN). 
\subsection{Feature Extraction}
PixelShuffle and wavelet transform have been proposed to replace pooling and interpolation to avoid information loss. Multi-level wavelet transform is considered by DSWN~\cite{2020Densely} to achieve better receptive field size and avoid down-sampling information loss by embedding wavelet transform into CNN architecture. DSWN owns more power to model both spatial context and inter-subband dependency by embedding DWT and IDWT to CNN. In this paper,our proposed network adopts the same method as DSWN to extract multi-scale information with a totally different architecture from DSWN.

As we know, stacking multiple convolutional layers can help us effectively extract high-dimensional information in the feature map. However, this will lead to a substantial increase in network parameters, requiring the massive amount of training data to prevent overfitting. In addition, the size of the convolution kernel used in the convolution layer will bring different receptive fields to feature extraction. The larger the size of the convolution kernel used, the larger the receptive field. But it will bring heavy computation and slower running time. With Densely Connected Residual Block, DRNet~\cite{2018Image} mitigates the problems of overfitting, vanishing gradient, and training instability during training very deep and wide networks. Moreover, it can improve the propagation and reuse of features by creating direct connections from the previous layers to the subsequent layers. Recently, Injecting holes in the standard convolution map, The dilated convolutions support exponential expansion of the receptive field without loss of resolution or coverage. The number of parameters associated with each layer is identical. The receptive field grows exponentially. Utilizing the dilation convolutions and Self-guided leanring strategy, MLGRB~\cite{WANG2019206} can extract the spatial contextual conformation to enhance further the feature representation ability. Inspired by MLGRB, we adopts the similar architecture as MLGRB to further acquire multi-scale information in different resolution space.
\begin{figure*}[t]
  \centering
   \includegraphics[width=1.0\linewidth]{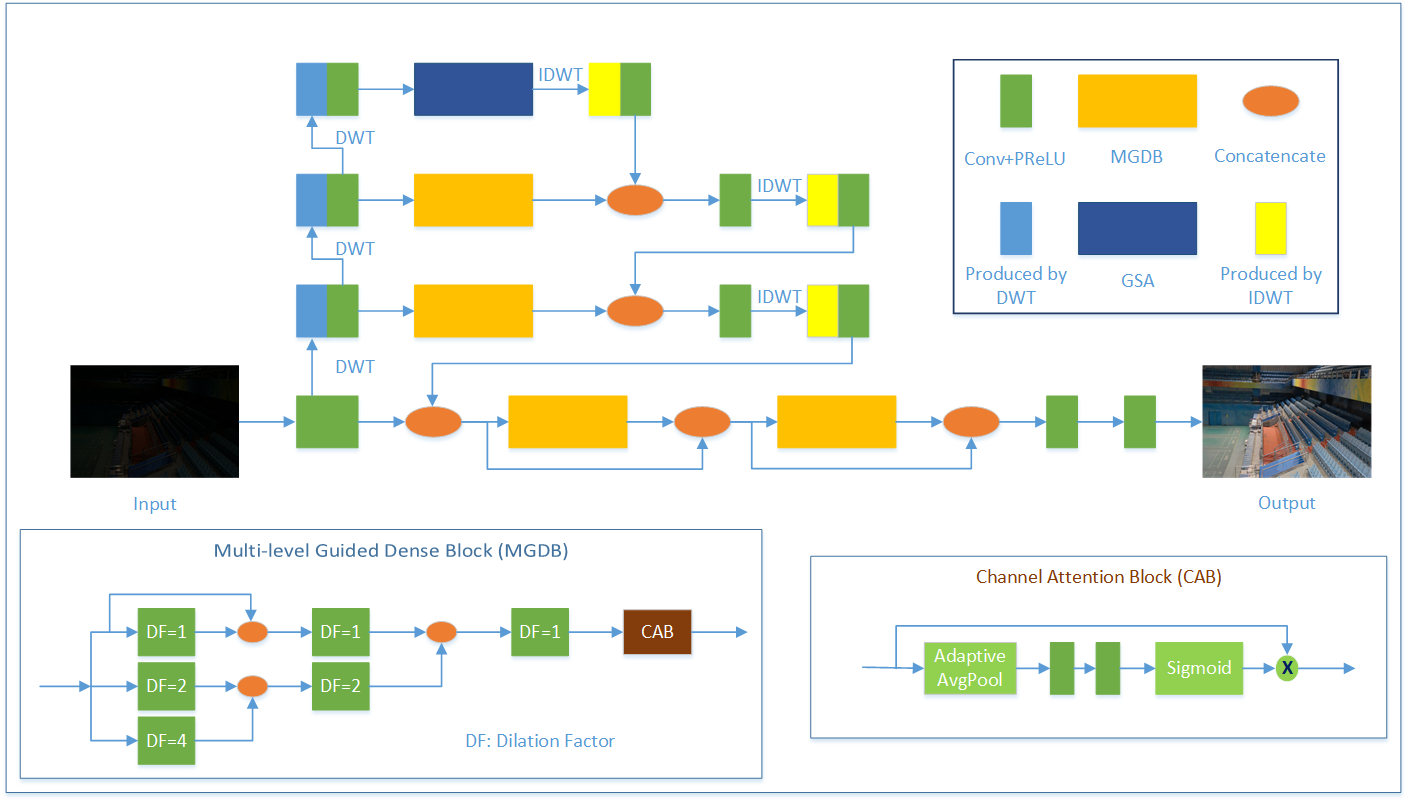}
   \caption{An illustration of our proposed network}
\end{figure*}
\section{Attention based Broadly Self-guided  Network}
In this section, we firstly introduce the overall network structure and then introduce the details of ABSGN.
\subsection{Overall Structure of ABSGN}
Our proposed network use a top-down self-guidance architecture to better exploit image multi-scale information. Information extracted at low resolution is gradually propagated into the higher resolution sub-networks to guide the feature extraction processes. Firstly, we pass the input image through a Conv+PReLU layers to obtain a main feature map with 32 channels. Then instead of PixelShuffle and PixelUnShuffle,DWT and IDWT are used to generate multi-scale inputs. ABSGN uses wavelet transform to transform the main feature map to three smaller scales. At the lowest resolution layer, we offer a Global Spatial Attention (GSA) Block including spatial attention module~\cite{2018CBAM} to be aware of the global context/color
information. Apart from the lowest solution level, we add Multi-level Guided Densely Block (MGDB) one or two blocks  as shown in Figure 2. At the full resolution layer, we adopt Dense connetions consists of two MGDB to imporove the reuse of the main feature map. In the rest of this paper, we simply refer to the advantages of GSA and MGDB. we also make a comparative experiment to prove our conclusion. In addition, we ﬁnd batch normalization is harmful for the denoising performance and only use one normalization layer in this network. 
\subsection{Detail Structure of ABSGN}
different form DSWN, we fistly acquire the main feature map with 32 channels from the input image by a Conv+PReLU layer. the aim of which is to increase the feature map by increasing the number of channel. we use DWT and a Conv+PReLU layer to finish the downsampling process. Using a Conv+PReLU layer to decrease the number of channels,which not only can decrease the calculation but also fine-tune the down-sampled feature map, Performing the above downsampling operation three times in sequence to obtain feature maps of different sizes.

The top level of ABSGN works on the smallest spatial resolution to extract large scale information.As is shown in Figure 3, the top sub-network use the GSA block to get the global information. which contains two Conv+PReLU layers and AdaptiveAvgPool2d, AdaptiveMaxPool2d, interpolation (the Resize block in Figure 3) and Spatial Attention module (SPA). Particulary, given an input feature map, i.e.X, with a size of $H\times W\times C$, AdaptiveAvgPool2d and AdaptiveMaxPool2d is employed to extract the representative information. Average of them is used for the global information producing a feature map with a size of $1\times 1\times C$. Then, an interpolation function is utilized to upscale the feature map with global information which is processed by a Conv+PReLU to shrink the number of channels, yielding a global feature map with a size of $H\times W\times C_1$. 
\begin{figure}[t]
   \centering
   \includegraphics[width=1.0\linewidth]{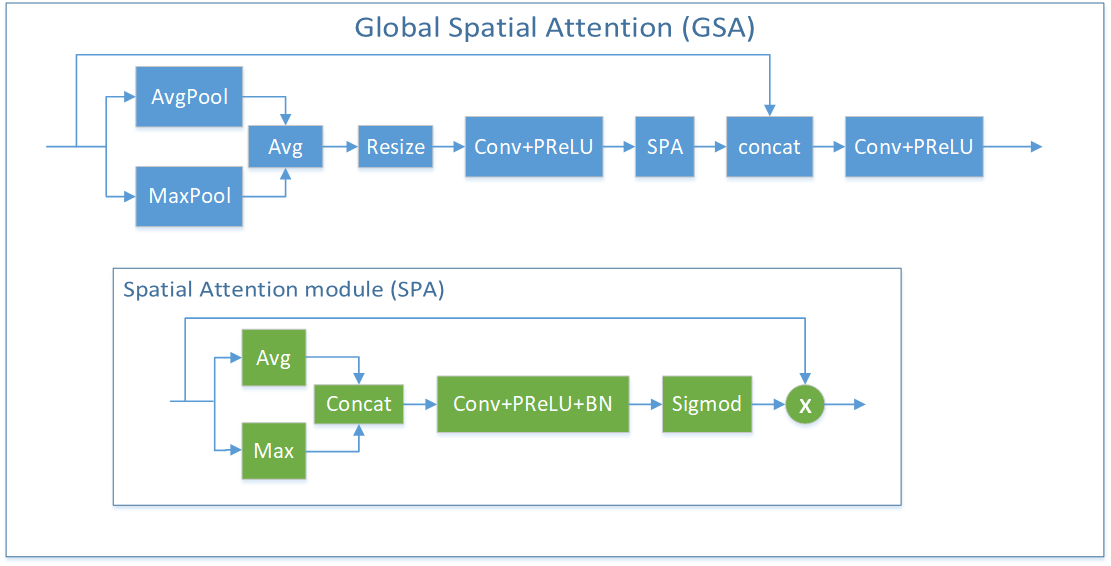}
   \caption{Diagram of Global Spatial Attention (GSA) block}
\end{figure} 
Then we apply SPA block to increase attention to different areas of the global feature map. The SPA block simultaneously applies max-pooling and average-pool Along channel dimention, then cancatenate the two feature maps to generate feature descriptors, the purpose of which is to highlight the information area. the feature descriptors then generates a spatial attention map through a convolution layer, Finally the map normalized by Activation function (Sigmoid) multiplies the input to obtain the output as optimized global feature map. Finally a concat function and a Conv+PReLU function are employed to combine the input feature map (encoding local information) and the optimized global feature map (encoding global information) to produce an output feature map with $H\times W\times C$.

Corresponds to the above downsampling process, we use IDWT and a Conv+PReLU layer to finish the upsampling process. At the middle two levels, 1x1 convolutional kernel layers are used to merge information extracted from different resolution. The network structure of the middle sub-networks take MGDB to fully extract the feature inforamation, which embed dilation convolution with different dilation factor into Dense Block. As is shown in Figure 2, we apply self-guided strategy to use the dilation covolution with greater dilation factor to guide the dilation covolution with the smaller one. before the final output, we add Channel Attention module~\cite{2018CBAM} to enhance the attention to different channels.

As for the full resolution level, we add more MGDB to reuse the main feature map, which enhance the feature extraction capability of ABSGN, after merging information from all the scales, we use two Conv+PReLU layers to acqiure the final output as the restored image. By adding gradient loss, our network is able to achieve better retention of details without reducing PSNR. Inspired by a new joint loss function~\cite{2015Loss}, our network uses L1 loss plus SSIM loss for training. The total loss is as follows:
\begin{equation}
\mit L_{ABSGN} = \gamma\mit L_{SSIM}(I,\hat{I})+(1-\gamma)\mit L_{l1}(I,\hat{I})
% MSE=\frac{1}{mn}\sum_{i=0}^{m-1}\sum_{j=0}^{n-1}[I(i,j)-K(i,j)]^{2}
\end{equation}
where $\gamma \in [0,1]$ is the weight to balance the two terms. Here we choose the value of the $\gamma$ is 0.16.
\section{Experiment}
In this section, we ﬁrst introduce the training details and provide experimental results on different datasets. Then we compare ABSGN with several state-of-the-art low-light image enhancement approaches.  
\begin{figure*}[t]
 \centering
 \begin{subfigure}{0.2\linewidth} 
 \centering
 \includegraphics[width=1.3in]{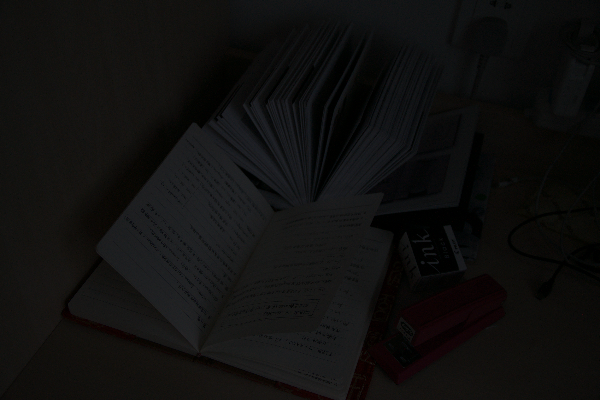}
 \caption*{(1) Input}
 \end{subfigure}\hspace{-2mm}
 \begin{subfigure}{0.2\linewidth} 
  \centering
 \includegraphics[width=1.3in]{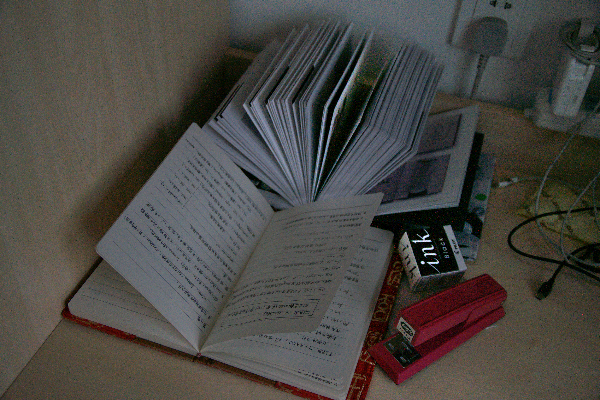}
 \caption*{(2) LIME}
 \end{subfigure}\hspace{-2mm}
  \begin{subfigure}{0.2\linewidth} 
   \centering
 \includegraphics[width=1.3in]{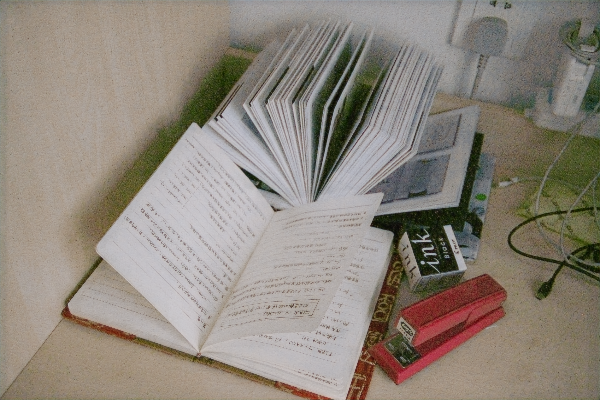}
 \caption*{(3) GLADNet}
 \end{subfigure}\hspace{-2mm}
  \begin{subfigure}{0.2\linewidth} 
   \centering
 \includegraphics[width=1.3in]{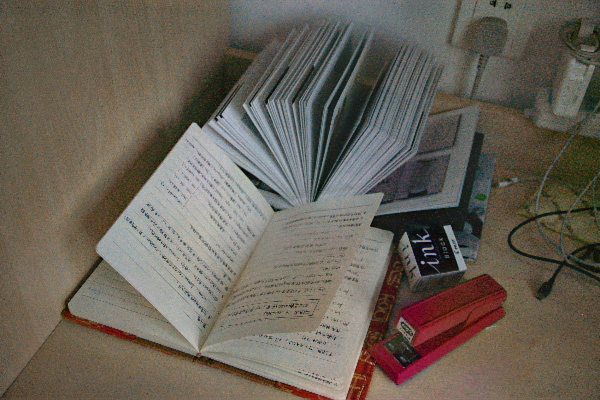}
 \caption*{(4) EnlightenGAN}
 \end{subfigure}\hspace{-2mm}
  \begin{subfigure}{0.2\linewidth}
  \centering
 \includegraphics[width=1.3in]{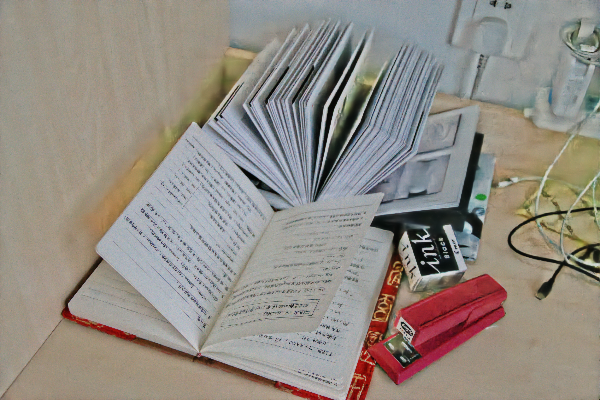}
 \caption*{(5) KinD++}
 \end{subfigure}
 
 \begin{subfigure}{0.2\linewidth} 
 \centering
 \includegraphics[width=1.3in]{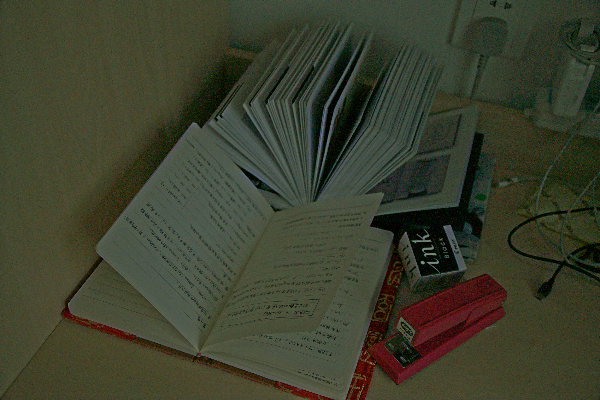}
 \caption*{(6) Zero-DCE}
 \end{subfigure}\hspace{-2mm}
 \begin{subfigure}{0.2\linewidth} 
  \centering
 \includegraphics[width=1.3in]{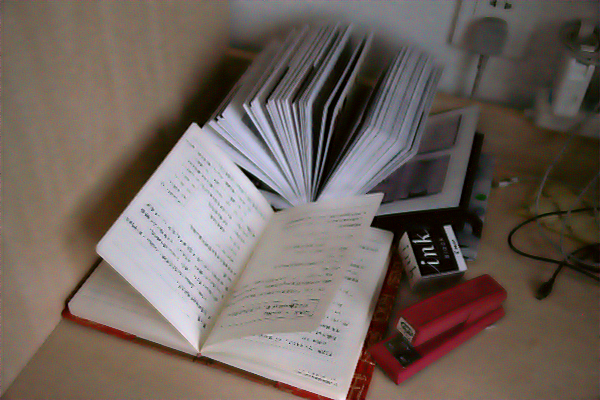}
 \caption*{(7) RUAS}
 \end{subfigure}\hspace{-2mm}
  \begin{subfigure}{0.2\linewidth} 
   \centering
 \includegraphics[width=1.3in]{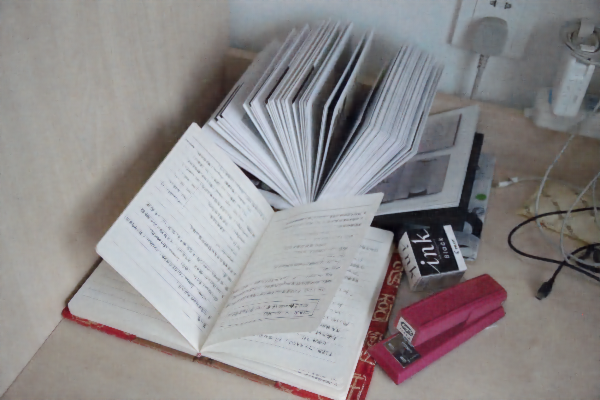}
 \caption*{(8) MIRNet}
 \end{subfigure}\hspace{-2mm}
  \begin{subfigure}{0.2\linewidth} 
   \centering
 \includegraphics[width=1.3in]{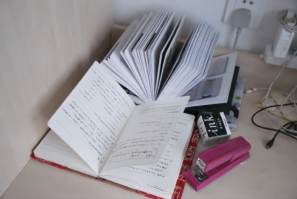}
 \caption*{(9) Ours}
 \end{subfigure}\hspace{-2mm}
  \begin{subfigure}{0.2\linewidth}
  \centering
 \includegraphics[width=1.3in]{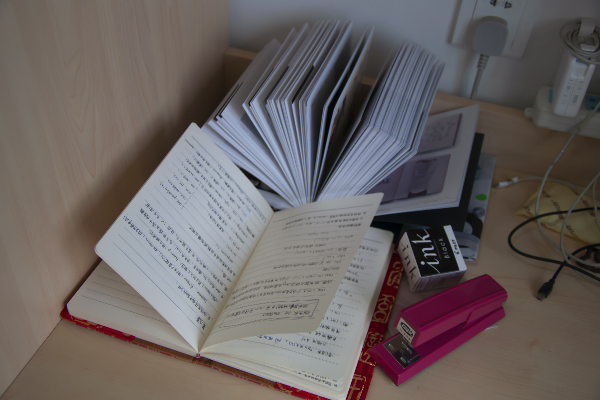}
 \caption*{(10) Reference}
 \end{subfigure}
\caption{Restoring results of the conventional methods and the proposed method on LOL dataset(lighter)}
\end{figure*}

\begin{figure*}[t]
 \centering
 \begin{subfigure}{0.2\linewidth} 
 \centering
 \includegraphics[width=1.3in]{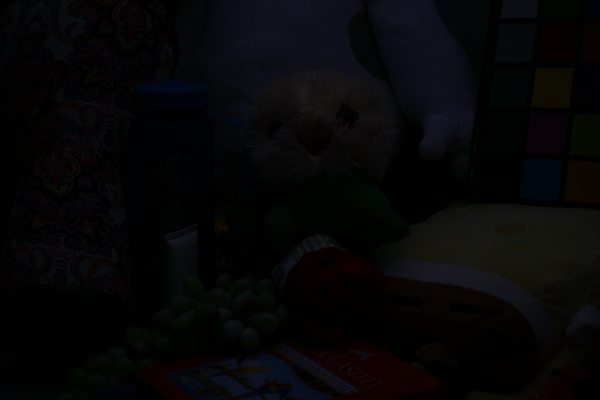}
 \caption*{(1) Input}
 \end{subfigure}\hspace{-2mm}
 \begin{subfigure}{0.2\linewidth} 
  \centering
 \includegraphics[width=1.3in]{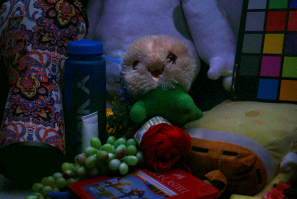}
 \caption*{(2) LIME}
 \end{subfigure}\hspace{-2mm}
  \begin{subfigure}{0.2\linewidth} 
   \centering
 \includegraphics[width=1.3in]{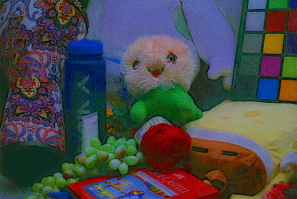}
 \caption*{(3) GLADNet}
 \end{subfigure}\hspace{-2mm}
  \begin{subfigure}{0.2\linewidth} 
   \centering
 \includegraphics[width=1.3in]{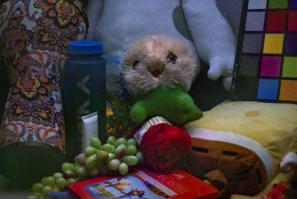}
 \caption*{(4) EnlightenGAN}
 \end{subfigure}\hspace{-2mm}
  \begin{subfigure}{0.2\linewidth}
  \centering
 \includegraphics[width=1.3in]{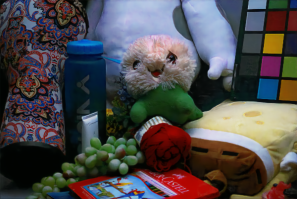}
 \caption*{(5) KinD++}
 \end{subfigure}
 
 \begin{subfigure}{0.2\linewidth} 
 \centering
 \includegraphics[width=1.3in]{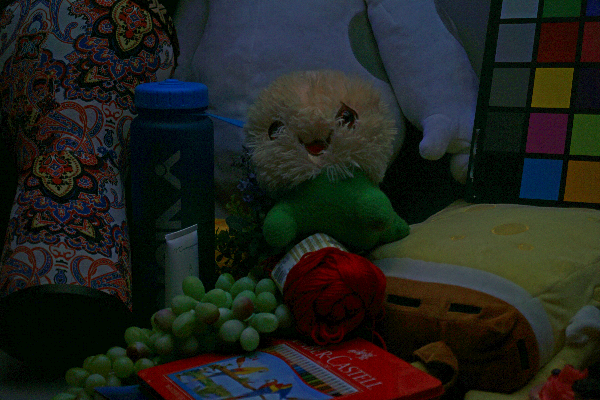}
 \caption*{(6) Zero-DCE}
 \end{subfigure}\hspace{-2mm}
 \begin{subfigure}{0.2\linewidth} 
  \centering
 \includegraphics[width=1.3in]{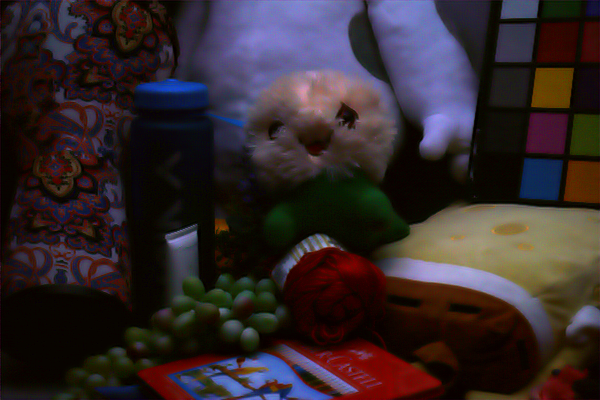}
 \caption*{(7) RUAS}
 \end{subfigure}\hspace{-2mm}
  \begin{subfigure}{0.2\linewidth} 
   \centering
 \includegraphics[width=1.3in]{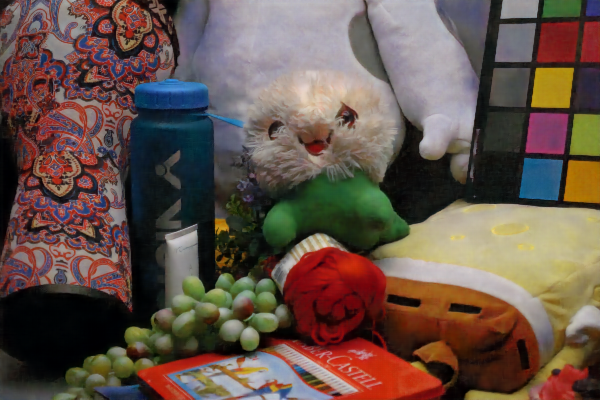}
 \caption*{(8) MIRNet}
 \end{subfigure}\hspace{-2mm}
  \begin{subfigure}{0.2\linewidth} 
   \centering
 \includegraphics[width=1.3in]{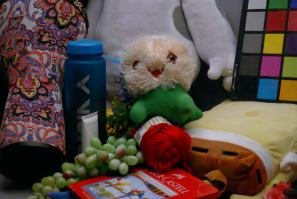}
 \caption*{(9) Ours}
 \end{subfigure}\hspace{-2mm}
  \begin{subfigure}{0.2\linewidth}
  \centering
 \includegraphics[width=1.3in]{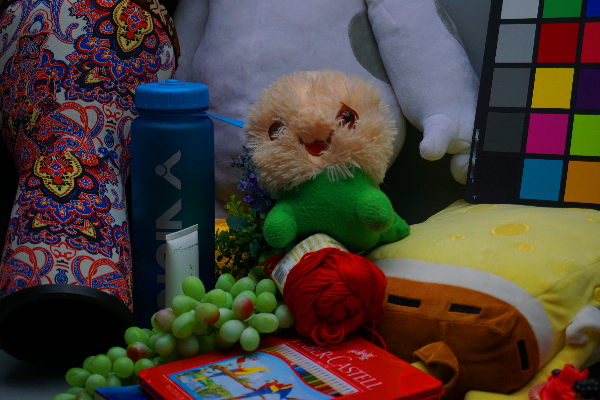}
 \caption*{(10) Reference}
 \end{subfigure}
\caption{Restoring results of the conventional methods and the proposed method on LOL dataset(Darker)}
\end{figure*}
\subsection{Experimental Setting}
Low-Light (LOL) dataset~\cite{wei2018deep} is a publicly available dark-light paired images dataset in the real sense. The low-light images are collected by changing exposure time and ISO. It contains 500 images in total, The resolution of each of these images is 400 × 600. we use 485 images of them for training, and the rest for evaluation as suggested by Most state-of-the-art low-light image enhancement solutions select the LOL as their training dataset~\cite{2018GLADNet,zhang2019kindling}. To train our ABSGN model, we also use LOL dataset as training and validation dataset for low-light image enhancement task. The LOL dataset consists of two parts: the real-world dataset and the synthetic dataset. Because the LOL synthetic dataset cannot simulate the degradation of real-world images very well, we deprecate the synthetic dataset where low-light images are synthesized from normal-light images and only use the real-world dataset. Our datasets for comparision also includes LIME~\cite{guo2016lime}, MEF~\cite{2015Perceptual}, DICM~\cite{2013Contrast} which are used by some recent low-light image enhancment networks~\cite{zhang2019kindling}. For these datasets lack of reference images, we use non-reference metric including No-reference Image Quality and Uncertainly Evaluator (UNIQUE)~\cite{zhang2021uncertainty} and Blind/Referenceless Image Spatial Quality Evaluator (BRISQUE)~\cite{2012No}. In addition, we also choose another commonly used data set called MIT-Adobe FiveK dataset for training and evaluation. MIT-Adobe FiveK~\cite{2011Learning} contains 5000 images of various indoor and outdoor scenes captured with DSLR cameras in different lighting conditions. The tonal attributes of all images are manually adjusted by five different trained photographers (labelled as experts A to E). we choose the enhanced images of expert C as the ground-truth. Moreover, the first 4500 images are used for training and the last 500 for testing. When training our model, we randomly crop 256 × 256 patches from the training images. The input patches of the proposed network are randomly flipped and rotated for data augmentation. The parameters of network are Kaiming initialized~\cite{2015Delving}. We train the whole network for 300 epochs overall. The learning rate is initialized as $1\times10^{-4}$ at the ﬁrst 200 epochs and reduce to $5\times10^{-5}$ in the next 50 epochs. We ﬁnetune our model at the last 50 epochs with a $1\times10^{-5}$ learning rate. For optimization, we use Adam optimizer ~\cite{2014Adam} with $\beta{1} = 0.5$,$\beta2 = 0.999$ and batch size equals to 5. We use L1 loss and SSIM loss for the total loss function, L1 loss is a PSNR-oriented optimization in the training process ~\cite{2015Loss}. SSIM loss can keep the overall structure. We adopt PSNR,SSIM~\cite{2004Image}, LPIPS ~\cite{zhang2018unreasonable} and FSIM ~\cite{zhang2011fsim} as the quantative metrics to measure the performance of our method. We use the Pytorch framework to build our model on the platform with Nvidia TITAN XP GPU and an Intel(R) Xeon(R) E5-2620 v4 2.10GHz CPU.

\subsection{Evaluation on LOL Dataset}
We compare our proposed network with several state-of-the-art low-light image enhancement solutions: MSRCR~\cite{Jobson1997A}, LIME, NPE, JED~\cite{2018Joint}, MBLLEN~\cite{lv2018mbllen}, RetinexNet, GLADNet, RDGAN, KinD++, Zero-DCE, EnlightenGAN and MIRNet. To compare the performance, we determined to take above measures as the objective and subjective measurements.
\begin{table}[t]\small
\centering
\caption{quantitative comparision of serveral metric between our method and state-of-the-art methods on LOL dataset.}
\begin{tabular}{c|cccc}
\hline
  \textbf{Method} & \textbf{SSIM}$\uparrow$ & \textbf{PSNR}$\uparrow$ & \textbf{LPIPS}$\downarrow$  &\textbf{FSIM}$\uparrow$   \\ \hline
    Input                     & 0.1914             & 7.77             & 0.4173            &0.7190      \\\hline
    MSRCR                     & 0.4615             & 13.17            & 0.4404            &0.8450         \\\hline
    LIME                      & 0.4449             & 16.76            & 0.4183            &0.8549       \\\hline
    NPE                       & 0.4839             & 16.97            & 0.4156            &0.8964    \\\hline
    JED                       & 0.6509             & 13.69            & 0.3549            &0.8812       \\\hline
    MBLLEN                    & 0.7247             & 17.86            & 0.3672            &0.9262           \\\hline
    RetinexNet                & 0.4249             & 16.77            & 0.4670            &0.8642           \\\hline
    GLADNet                      & 0.6820             & 19.72            & 0.3994            &0.9329           \\\hline
    RDGAN                     & 0.6357             & 15.94            & 0.3985            &0.9276           \\\hline
    KinD++                    & 0.8203             & 21.30            & 0.1614            &0.9424           \\\hline
    Zero-DCE                  & 0.5623             & 14.87            & 0.3852            &0.9276           \\\hline
    EnlightenGAN              & 0.6515             & 17.48            & 0.3903            &0.9226          \\\hline
    MIRNet                    & 0.8321             & 24.14            & 0.0846            &0.9547           \\\hline
 \textbf{ours} & { \textbf{0.8680}} & { \textbf{24.59}} & {\textbf{0.0772}} & {\textbf{0.9659}} \\ \hline
\end{tabular}
\end{table}
\begin{table}[ht]\scriptsize%\footnotesize
\caption{$UNIQUE\uparrow/BRISQUE\uparrow$ Comparison on NPE, LIME, MEF, DICM}
\centering
\renewcommand{\arraystretch}{2.0}
\begin{tabular}{c|ccc|c}
\hline
\textbf{Method}      &       \textbf{LIME}   &     \textbf{MEF}   &   \textbf{DICM}  &\textbf{Average}\\ [3pt]\hline
    Dark             & 0.826/21.81       & 0.738/23.56       &0.795/21.57    &0.786/22.31  \\[3pt]
    PIE              & 0.791/22.72       & 0.752/11.02       &0.791/21.72    &0.778/18.49   \\[3pt]
    LIME            & 0.774/20.44       & 0.722/15.25       &0.758/23.48    &0.751/19.72  \\[3pt]
    MBLLEN          & 0.768/30.26       & 0.717/37.44       &0.787/32.44    &0.757/33.38     \\[3pt]
    RetinexNet       & 0.794/31.47       & 0.755/20.08       &0.770/29.53    &0.773/27.03   \\[3pt]
    KinD            & 0.766/39.29       & 0.747/31.36       &0.776/32.71    &0.763/34.45   \\[3pt]
    Zero-DCE         & 0.811/21.40       & 0.762/16.84       &0.777/27.35    &0.783/21.86    \\[3pt]
    MIRNet           & 0.814/33.73       & 0.768/21.45       &0.812/33.71    &0.798/29.63   \\[3pt]\hline
 \textbf{ours}  & { \textbf{0.827 / 32.23}} & { \textbf{0.784 / 38.52}} & { \textbf{0.804 / 33.23}}  & { \textbf{0.806 / 34.66}}\\[3pt] \hline
\end{tabular}
\vspace{-2mm}
\end{table}
As shown in table 1, our method achieves the best performance in PSNR, SSIM, LPIPS and FSIM, MIRNet is the second best method, Although MIRNet adpots more complicated structure and has more than ten times the number of parameters, ABSGN is still able to achieve better PSNR and SSIM than MIRNet on average. Figure 4 and Figure 5 show some example results in LOL dataset with different light levels. we can see that our ABSGN shows the best PSNR and SSIM in different light levels. There also show some detail results of several low-light image enhancement, where KinD++ is a classical decomposing network and MIRNet has excellent feature extraction capabilities. We can see all the three low-light image enhancment networks are able to achieve a obvious improvement compared with low light images. ABSGN is better than KinD++ and MIRNet in some details such as texture details of the book in Figure 4 and the color of the embroidery in Figure 5. Our proposed method is able to handle different light levels and reserve more details at the same time.
\begin{figure*}[t]
\centering
 \begin{subfigure}{0.24\linewidth}
  \begin{minipage}[t]{0.24\linewidth}
	\includegraphics[width=4cm,height=3cm]{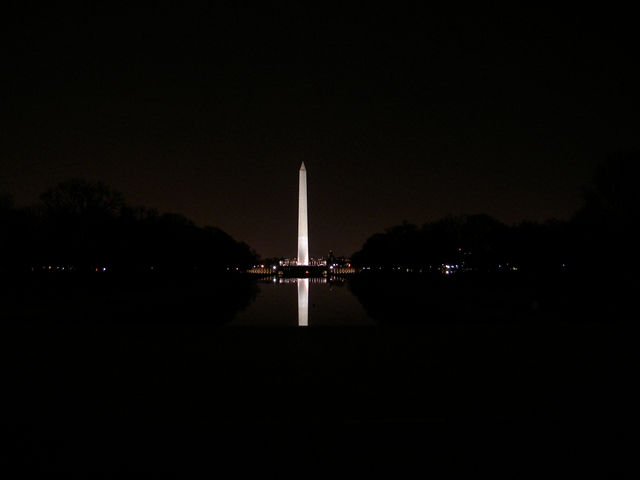}\vspace{0.1cm}
	\includegraphics[width=4cm,height=3cm]{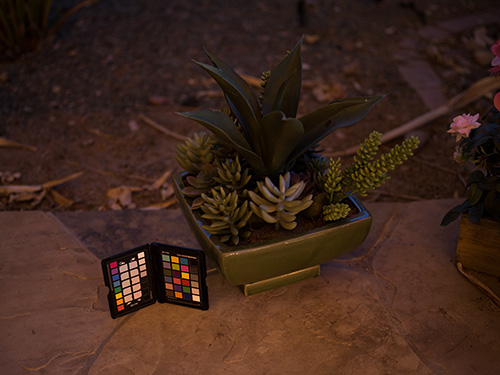}\vspace{0.1cm}
	\includegraphics[width=4cm,height=3cm]{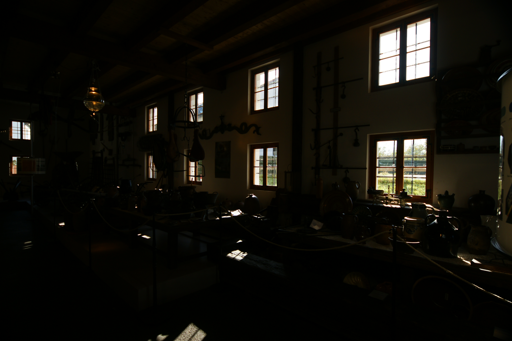}
 	\end{minipage}	
 	\caption*{Input}
\end{subfigure}\hspace{-0.1cm}
\begin{subfigure}{0.24\linewidth}
\begin{minipage}[t]{0.24\linewidth}  
\includegraphics[width=4cm,height=3cm]{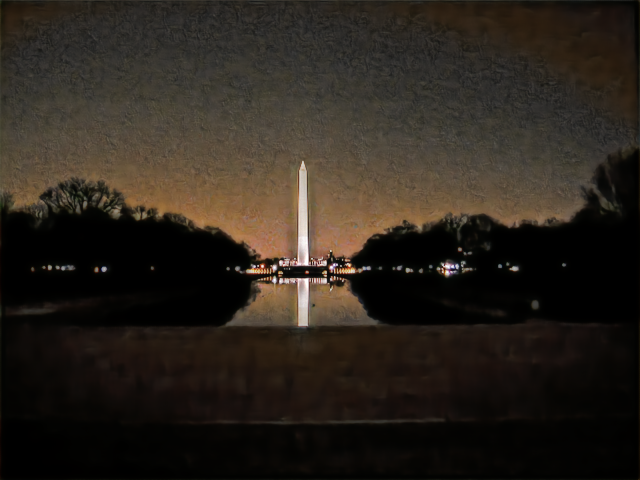}\vspace{0.1cm}
\includegraphics[width=4cm,height=3cm]{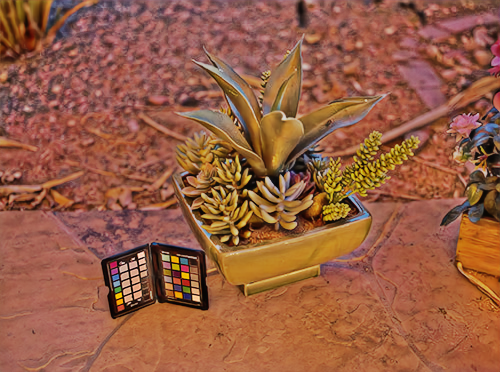}\vspace{0.1cm}
\includegraphics[width=4cm,height=3cm]{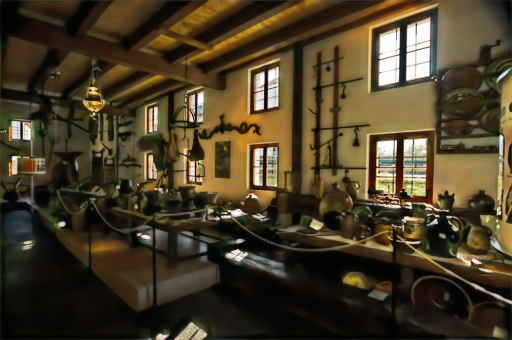}
\end{minipage}
\caption*{KinD++}
\end{subfigure}\hspace{-0.1cm}
\begin{subfigure}{0.24\linewidth}
\begin{minipage}[t]{0.24\linewidth} 
\includegraphics[width=4cm,height=3cm]{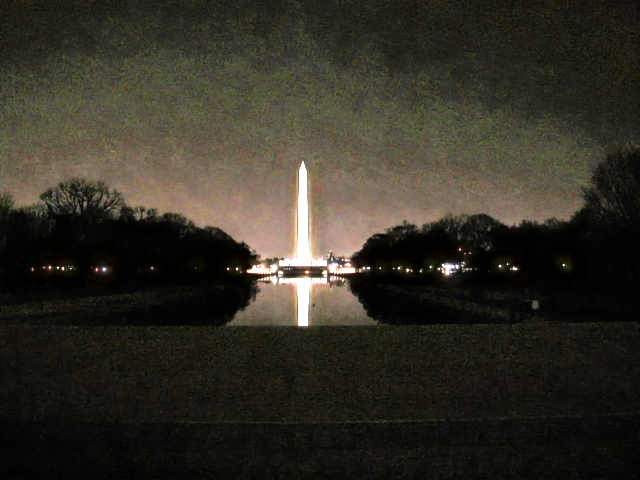}\vspace{0.1cm}
\includegraphics[width=4cm,height=3cm]{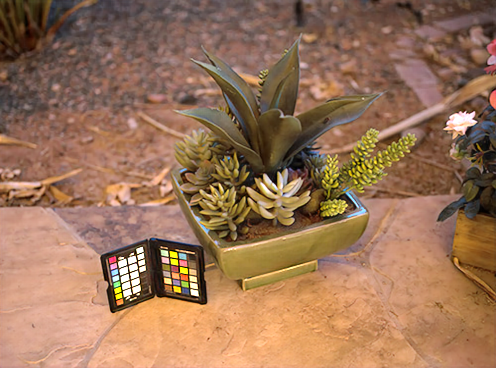}\vspace{0.1cm}
\includegraphics[width=4cm,height=3cm]{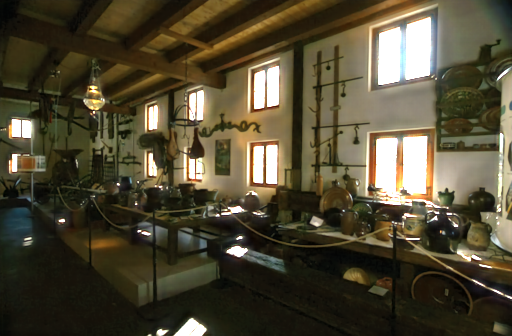}
\end{minipage}
\caption*{MIRNet}
\end{subfigure}\hspace{-0.1cm}
\begin{subfigure}{0.24\linewidth}
\begin{minipage}[t]{0.24\linewidth} 
\includegraphics[width=4cm,height=3cm]{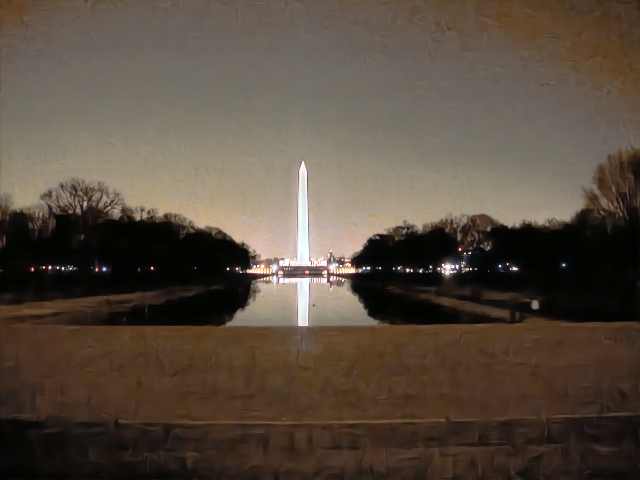}\vspace{0.1cm}
\includegraphics[width=4cm,height=3cm]{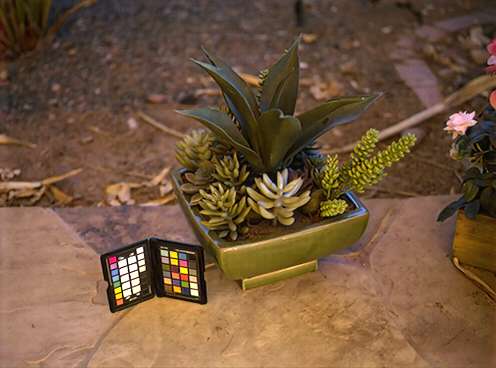}\vspace{0.1cm}
\includegraphics[width=4cm,height=3cm]{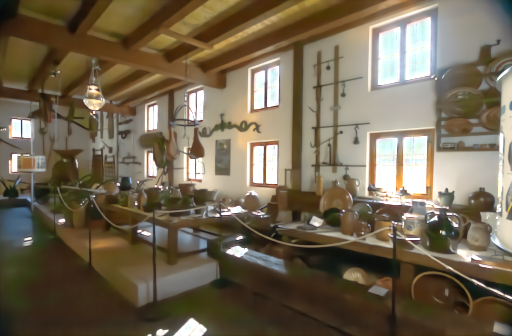}
\end{minipage}
\caption*{Ours}
\end{subfigure}
\caption{Visual Comparision on DICM,LIME,MEF from top to bottom. from left to right:Dark,KInD++,MIRNet,ours }
\vspace{-0.2cm}
\end{figure*}
\begin{figure*}[t]
 \centering
 \begin{subfigure}{0.33\linewidth} 
 \centering
 \includegraphics[width=1.5in]{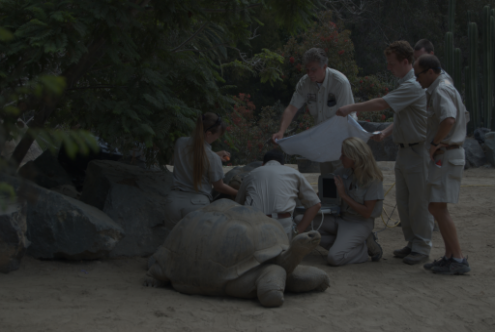}
 \caption*{(1) Input}
 \end{subfigure}\hspace{-2mm}
 \begin{subfigure}{0.33\linewidth} 
  \centering
 \includegraphics[width=1.5in]{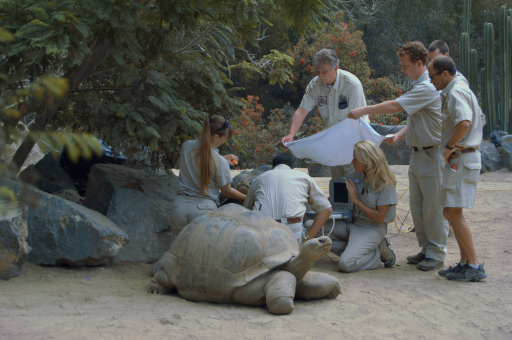}
 \caption*{(2) HDRNet~\cite{gharbi2017deep}}
 \end{subfigure}\hspace{-2mm}
  \begin{subfigure}{0.33\linewidth} 
   \centering
 \includegraphics[width=1.5in]{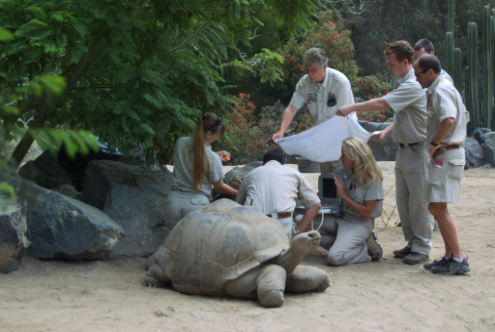}
 \caption*{(3) DeepUPE~\cite{2019Underexposed}}
 \end{subfigure}
  
 \begin{subfigure}{0.33\linewidth} 
 \centering
 \includegraphics[width=1.5in]{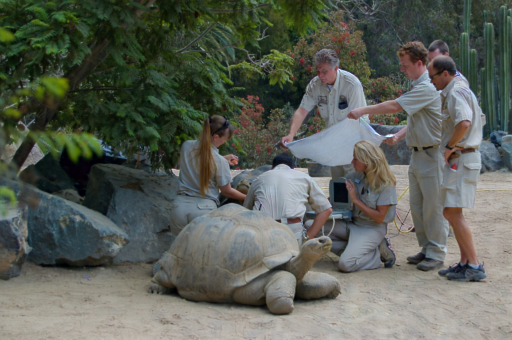}
 \caption*{(4) MIRNet}
 \end{subfigure}\hspace{-2mm}
 \begin{subfigure}{0.33\linewidth} 
  \centering
 \includegraphics[width=1.5in]{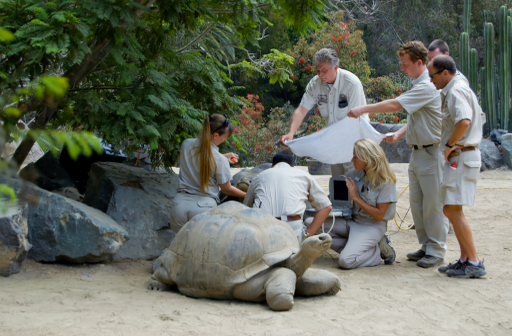}
 \caption*{(5) ABSGN(ours)}
 \end{subfigure}\hspace{-2mm}
  \begin{subfigure}{0.33\linewidth} 
   \centering
 \includegraphics[width=1.5in]{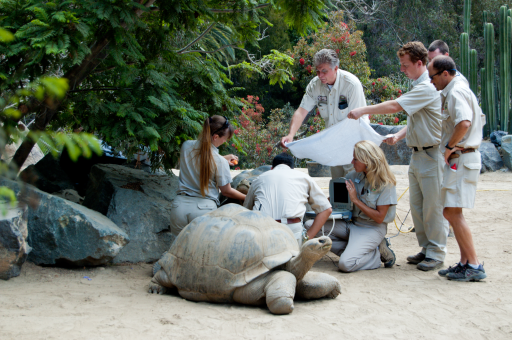}
 \caption*{(6) References}
 \end{subfigure}
\caption{Restoring results of the conventional methods and the proposed method on MIT-Adobe FiveK dataset}
\end{figure*}
Table 2 shows the comparision on LIME, MEF and DICM datasets.Our model has the best average UNIQUE and the best average BRISQUE. From figure 6, we can also conclude a similar conclusion to the LIME, MEF and DICM datasets. At a darker region, Restoring networks tend to smooth the noise too much, because the network is difficult to distinguish true details from the dark region. ABSGN can better preserve details at dark regions,such as soils and floors.

\subsection{Evaluation on MIT-Adobe FiveK dataset}
We report PSNR/SSIM values of our method and several other techniques in Table 3 for the MIT-Adobe FiveK datasets,respectively. It can be seen that our Network achieves significant improvements over previous approaches.Notably,when compared to the recent best methods, ABSGN obtains 1.52 dB performance gain over over MIRNet on the Adobe-Fivek dataset.
\begin{table}[t]\footnotesize
\caption{Image Enhancement between our method and state-of-the-art methods on MIT-Adobe FiveK dataset.}
\centering
\renewcommand{\arraystretch}{2.0}
\begin{tabular}{cccccc}
\hline
  \textbf{Method}      & \textbf{HDRNet}   &   \textbf{DPE} & \textbf{DeepUPE}  &\textbf{MIRNet} &\textbf{ours}\\ \hline
    PSNR               & 21.96           &22.15    &23.04  &23.73  &25.25  \\
    SSIM               & 0.866           &0.850    &0.893  &0.925  &0.931    \\\hline
\end{tabular}
\vspace{-3mm}
\end{table}
We also show visual results in Figure 7. Compared to other techniques, our method generates enhanced images that are natural and vivid in appearance and have better global and local contrast.
\subsection{Ablation study}
In this part, to prove the effectiveness of the proposed module and the neccessity of introducing the method, we have made comparative Experiment. besides, we did another comparative experiment to emphasize the advantage and practical value of ABSGN. 
\subsubsection{Contribution of our proposed module}
This ablation study is to answer the question that why not just adpot Dense Connetly Residual (DCR) Block to get the local information and using Global Attention Aware (GIA) module like GIANet~\cite{meng2020gia} to get the global information like many end-to-end Network, if the network adopt the conventional Global Attention module,Whether there will be better effect. As is mentioned in Sec.3.1, the MGDB adpot the similar self-guided strategy to introduce the dilation convolution,which has larger receptive ﬁeld and better local feature extraction capacity. the CNN based method GSA module is introduced in our proposed method to avoid the problems that too long convolution layers cause the loss of global information.

In Table 4, we show the ablation study of these two modules.we use the DenseRes Block to replace our MGDB to test the effect. In Addition, we try to apply the custom convolution to take place of all the dilation convolutions, which will tell our the necessity of introducing Dilation Convolution (DC). The results suggest that MGDB can get more local information and better performance in SSIM/PSNR. Similarity, we respectively use SPA module and GIA module substitute for our GSA module to prove the advantage of our proposed module. As is shown in Table 4, Compared to SPA module and GIA module,our module has irreplaceable advantage. these above module can’t replace our GSA module to extract the global information. 
\begin{table}[t]\small%\tiny%\footnotesize
\caption{Comparision experiment using different modules on LOL dataset.}
\centering
\setlength{\tabcolsep}{2mm}{%自动调整表格的宽度
\begin{tabular}{c|ccccc}
\hline
  \textbf{Method} & \textbf{SPA} &  \textbf{GIA} & \textbf{DCR}  &  \textbf{w/o DC}  & \textbf{ours} \\ [4pt]\hline
    PSNR(dB)        & 22.30           & 23.42        & 22.67           & 23.16              &24.59      \\
    SSIM            & 0.8456          & 0.8515       & 0.8483          & 0.8546             &0.8675       \\\hline
\end{tabular}}
\end{table}
\subsubsection{Adavantage of ABSGN}
We compare BSWN with other state-of-art deep learning low-light image enhancement methods in terms of the number of parameter,the inference time and UQI~\cite{wang2002universal} in LOL Dataset. As is shown in tabe 5, our model has the best UQI, meaning our results has closest information with the reference images.

\begin{table}[t]
\caption{Runtime cost and performance comparison of our mothod and other state-of-art deep learning methods on the LOL dataset.}
\centering
\begin{tabular}{c|cc|c}
\toprule
Deep learning Method & Params & Time cost & UQI$\uparrow$\\
\midrule
    MBLLEN      & 1.95M    & 80ms    & 0.8261 \\
    RetinexNet  & 9.2M     & 20ms    & 0.9110 \\
    GLAD        & 11M      & 25ms    & 0.9204 \\
    RDGAN       & 4.2M     & 30ms    & 0.8296 \\
    Zero-DCE    & 0.97M    & 2ms     & 0.7205 \\
    ElightenGAN & 33M      & 20ms    & 0.8499 \\
    KinD++      & 35.7M    & 280ms   & 0.9482 \\
    MIRNet      & 365M     & 340ms   & 0.9556 \\
    Our model   & 33M      & 14ms    & 0.9589 \\
\bottomrule
\end{tabular}
\label{tab:example}
\end{table}
\begin{figure}[ht]
  \centering
   \includegraphics[width=1.0\linewidth]{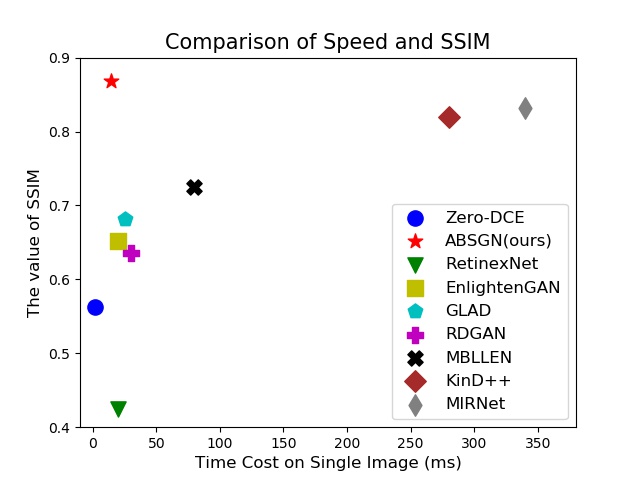}
   \caption{Runtime cost and SSIM comparison of our method and other state-of-the-art deep learning methods on the LOLdataset.}
\end{figure}
Although the parameters of our model is a little heavy, MIRNet is ten times bigger than ours. what’s more, inference speed of our model is very fast. As is shown in Figure 8, we tested all of the deep learing models on the same platform. The closer the method is to the upper left corner in the figure, the faster the model is and the higher the SSIM is. As is shown in the last colume of table in table 5. our model only take 14 ms to process a real-world low-light image with resolution of 600x400 from the LOL dataset. The speed of our model ranks second among all of the compared methods. Although Zero-DCE has very fast inference speed,which is only to deal with a little darker image. the biggest problem of Zero-DCE is difficult to restore the dark image especially in the night, which leads to its lack of strong applicability and application prospects. In order to reduce the depth of the network, make full use of the system's ability to parallelize, reduce the depth of the network by strengthening the width of the network, our model apply self-guidance strategy to parallel deal with the feature map in different resolutions. This is why our model can run at such a high speed keeping excellent restoring capacity.    

\section{Conclution}
In this paper, we proposed a Attention based Broadly Self-guided Network (ABSGN) for low-light image enhancement. ABSGN adopts a top-down manner to restore the low-light images. We use wavelet transform and a conv+PReLU layer to generate input variations with different spatial resolutions. At the lowest solution, we design GSA module to fully collect the global information. Then, we embed a MGDB into the middle two low spatial resolution levels to fully get local information. respectively, At the full resolution level, we employ more MGDB and a Dense connections to further reuse the information of the main feature map from the input image. The proposed ABSGN was validated on image restoration and real world low-light image enhancement benchmark and ABSGN is able to generate higher quality results than the compared state-of-the-art methods. Further, our ABSGN has excellent inference speed, which has good practical value and application prospects. 
%%%%%%%%% REFERENCES
{\small
\bibliographystyle{ieee_fullname}
\bibliography{egbib}
}

\end{document}